\documentclass[prd, twocolumn, nofootinbib, floatfix]{revtex4-2}
\usepackage{amsmath}
\usepackage{graphicx}
\usepackage{dcolumn}
\usepackage{bm}
\usepackage{epsfig}
\usepackage{amssymb,latexsym,mathrsfs}
\usepackage{graphicx}
\usepackage{color}
\usepackage{hyperref}

\usepackage{float}

\hypersetup{
    colorlinks=true,
    linkcolor=red,
    citecolor=blue,
} 
\newcommand{\be}{\begin{equation}}
\newcommand{\ee}{\end{equation}}
\newcommand{\bs}{\begin{split}} 
\newcommand{\bea}{\begin{eqnarray}}
\newcommand{\eea}{\end{eqnarray}}

\begin{document}

\title{Distinguishing Time Clustering of Astrophysical Bursts} 

\author{Mikhail Denissenya$^1$, Bruce Grossan$^{1,2}$, Eric V.\ Linder$^{1,2,3}$} 
\affiliation{${}^1$Energetic Cosmos Laboratory, Nazarbayev University, 
Nur-Sultan 010000, Kazakhstan\\
${}^2$Space Sciences Laboratory, 
University of California, Berkeley, CA 94720, USA\\
${}^3$Berkeley Center for Cosmological Physics \& Berkeley Lab, 
University of California, Berkeley, CA 94720, USA
}

\begin{abstract} 
Many astrophysical bursts can recur, and their time series structure or pattern 
could be closely tied to the emission and system physics. While analysis of 
periodic events is well established, some sources, e.g.\ some fast radio bursts and soft gamma-ray emitters, are suspected of more subtle and less explored periodic windowed behavior: the bursts 
themselves are not periodic, but the activity only occurs during periodic 
windows. We focus here on distinguishing 
periodic windowed behavior from merely clustered events through  
time clustering analysis, using techniques analogous to spatial clustering, 
demonstrating methods for identifying  
and characterizing the behavior. 
An important aspect is accounting for the ``curious 
incident of the dog in the night time'' -- lack of bursts carries 
information. 
As a worked example, we analyze six years of data from the soft gamma repeater 
SGR1935+2154, deriving a 
window period of 231 days and 55\% duty cycle; this has now successfully 
predicted both active and inactive periods. 
\end{abstract}

\date{\today} 

\maketitle

\section{Introduction}

The time evolution of 
energetic astrophysical burst events 
carries critical information and clues to their nature. This can 
reveal exciting, 
extreme astrophysics such as 
complete stellar disruption, ultrahigh magnetic fields, 
accompanying neutrino bursts, etc.\ and  
high luminosities visible to great distances. 
Notable high luminosity examples include 
gamma ray bursts (GRB), fast radio bursts (FRB), and soft gamma repeaters 
(SGR). 
Important puzzles remain concerning their nature, involving  
aspects of stellar structure, accretion, jet production, 
and the circumobject medium. 

Of particular interest are those events that recur, indicating that the 
process is not wholly disruptive, and possibly involves rotation 
or orbits; they also offer the possibility of observing 
the explosive event multiple times. Furthermore, 
if the repetition can be somewhat 
predicted, so that observations can be scheduled, this enables enhanced 
opportunities for understanding the burst mechanism and astrophysics. 
Orbits and rotation naturally impose periodic modulation on 
a variety of astrophysical signals, from occultations of 
stars by planets to pulsars to accretion disk phenomena. 
As a result there is a highly developed set of mathematical, statistical, and computer code tools to 
estimate the statistical likelihood of periodicity in a 
noise-limited or non-ideal sampling of data, and then 
determining the period and its uncertainty. 
This field has shown diverse evolution and activity from early Blackman-Tukey analysis 
\cite{blackmantukey} to ``pulsar folding'' 
(e.g.\ \cite{pulsarfolding}) to Lomb-Scargle periodograms 
(see, e.g., \cite{vdplas} for a review of the Lomb-Scargle periodogram and comparison to other methods). 

An example of greater complexity of phenomena 
and data came with the discovery and study of quasi-periodic 
oscillations -- such an X-ray emission phenomenon linked to 
accretion appears not as a simple periodic signal but as a  
broad bump in frequency space. 
Our focus here is 
periodic windowed behavior (PWB), inspired by the recent discovery of such behavior in repeating FRB sources \cite{cruces21,rajwade20}.  
For PWB, activity occurs only during periodically occurring windows  -- there is no activity in the gaps between the active windows, however not all active windows may show activity. The activity within a window may be random; there is no requirement or expectation that it will converge to a uniform profile (like a pulsar profile). 
Both the period and the active 
fraction (e.g.\ related to duty cycle of the energetic 
astrophysical process) are of interest. 

Radio telescopes observing repeating FRB sources report millisecond duration 
bursts from the same source that may have time spacings from 
milliseconds to days during continuous observations, but intensive
monitoring campaigns may observe no bursts 
for $\sim180$ days \cite{rajwade20,cruces21}. Initially, it was 
proposed that the behavior of the best-known repeating FRB, the source of FRB 121102,
may be modeled as a time-clustered 
Weibull distribution \cite{oppermann18}. However, with a few years of 
data in hand, an unexpected, and stunning, result appeared:
the bursts were not periodic, or simply clustered, but were 
observed only in``periodic activity windows'' \cite{rajwade20,cruces21}. 
Similar behavior was reported for the source of 
FRB 180916 \cite{chime20_180916}.  

More recently, PWB was reported in 
SGR1935+2154, but in soft gamma-ray bursts \cite{grossan20}.  This object is particularly noteworthy as it became the only known source of FRBs within our Galaxy when two were detected during the same day in 2020 \cite{kirsten+20,bochenek20}. This object is so much closer than any other known FRB source that one could hope it could be some kind of ``Rosetta stone'' for understanding FRBs. 
(Another claim of PWB in soft gamma bursts was given for SGR1806-20 \cite{2101.07923}, though it is 
less certain.) 

Robust identification, and characterization, of PWB could shed 
light on astrophysical burst mechanisms and energetics. For example, a 
considerable number of theories have been proposed for the origin of FRBs \cite{frbtheorycat}; 
documentation and analysis of PWB in FRBs would be very important in constraining these models and 
could be key to understanding the bursting nature. 
As we enter an age of 
big data time domain surveys, PWB could also be discovered in other 
astrophysical contexts, showing further value for improved ways 
of identifying and measuring this phenomenon. 

In Section~\ref{sec:id} we describe a method for identifying PWB, and in  
particular distinguishing it from more irregular time clustering of events. 
For a burst series suspected of having PWB, in Section~\ref{sec:det} 
we present methods for determining the period and active fraction. We  
apply this to the data from SGR1935+2154 in Section~\ref{sec:sgr}, and 
discuss extensions and conclude in Section~\ref{sec:concl}.

\section{Identifying Periodic Windowed Behavior} \label{sec:id} 

Strictly periodic behavior can be identified through a large number of 
different methods; for astrophysical time series data one of the most 
widely used is the Lomb-Scargle periodogram \cite{lomb,scargle,vdplas}, 
which 
works in frequency space, where the folding of the time series increases 
the signal. 
In periodic windowed behavior, activity occurs at some times 
(not necessarily deterministic) within 
windows, where it is the windows that recur at a regular period. 
Such behavior may be the result of, e.g., a periodic ``shutter'' modulating non-periodic emission, 
or the physical conditions necessary for an outburst 
may occur only periodically (but not guarantee a specific 
time for a burst), or see \cite{katz} and references therein 
for an FRB precessing beam model. 
In this case long-term folding of data will not 
necessarily converge to an average profile. 

The active windows may be a significant fraction of the full period, and then often 
the frequency analysis methods are diluted, with aliasing of the signal, i.e.\ signal peaks 
are broadened. 
Moreover, they tend to ignore the ``curious incident of the dog in the 
night time'' \cite{holmes}: that the dog didn't bark (the gaps with no 
bursts) carries important information. Therefore we discuss a time domain 
method for assessing the burst data. 

Working in the time domain, we initially 
examine the cumulative distribution function (CDF), 
since it does contain information on both events 
and gaps, and has useful statistical properties. 
From an ordered series of 
events at times $\{t_i\}$, we look at the distribution of burst spacings 
$t_i-t_1$ vs observation time. Normalizing both axes to run from 0 to 1 (at the 
end we can restore the scaling by the total 
observation time, to get answers in days), 
we have the cumulative fraction of events, between 0 and 100\%, vs the 
time fraction, normalized by the length of the observation 
campaign\footnote{Some 
technical details: we take time 0 to be when the first burst is observed; 
any gap before this is uninformative since we have no way of knowing whether 
the object had ever burst before -- basically we are interested in the 
repeats. The end time can be either the end of the observing  
or when the last burst is detected; this does not affect the results. 
We then normalize all burst time differences $t_i-t_1$ by 
$t_{\rm end}-t_1$ so that the $x$-axis of time fraction runs from 
[0,1].}. 
Thus we have the CDF of a variable distributed on [0,1]. 

\begin{figure}[htb!]
\centering 
\includegraphics[width=\columnwidth]{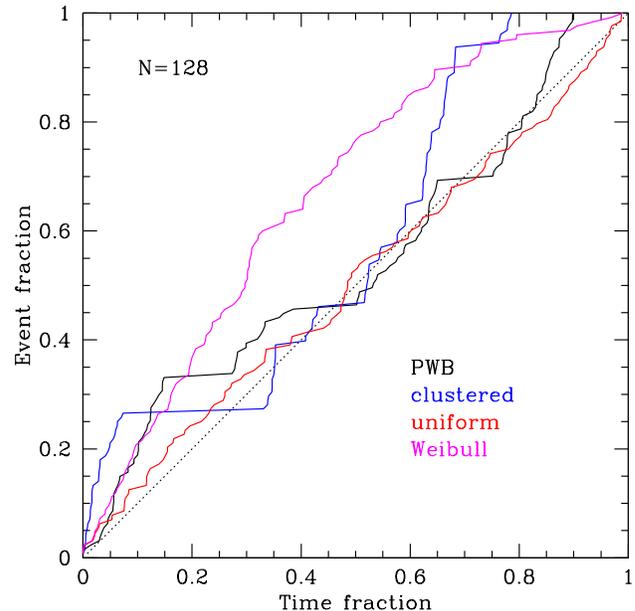}
\caption{Cumulative distribution functions are shown for random  
realizations of four different distributions, with a mean of 128 
events per distribution. The event fraction, 
number of events as a fraction of total number of events, is plotted 
vs the time fraction, how far into the total observing duration the 
event occurs. The dotted diagonal shows a perfect uniform distribution. 
The other distributions exhibit clusters of events with gaps of no 
activity interspersed. One can see at least hints of periodic windowed  
behavior in the PWB distribution. 
} 
\label{fig:pwball128} 
\end{figure}

Figure~\ref{fig:pwball128} illustrates several CDF for various realized 
distributions with a mean of 128 events each. We compare the uniform random 
distribution to two clustered distributions -- all without actual 
periodicity -- to a PWB case. The distributions have both similarities
and differences, and the eye can be fooled by random excursions into
thinking it detects patterns, even periodic ones.

The uniform distribution is realized by generating event times in a 
uniform random fashion over the [0,1] interval. The Weibull distribution 
is often used in astrophysics (e.g.\  \cite{oppermann18,cruces21} for FRBs), 
and more widely to give events that 
are clustered together at either early times or late times (as in failure 
rapidly or due to aging). 
We show an example with scale parameter 
$\beta=0.4$ and shape parameter $k=1.5$, so the CDF is 
$1-\exp\{-(t/\beta)^k\}$. Events beyond the end of the unit interval 
are not yet observed and not selected. The clustering here is due solely  
to the distribution. A true clustering is implemented in the clustered 
distribution, where events are generated with a two point correlation 
function $\xi(r\equiv |t_j-t_i|)\sim r^{-\gamma}$. We construct this 
using the Soneira-Peebles approach \cite{soneira,pjep80}, using a multiplicity 
parameter $\eta=4$, scale parameter $\lambda=8$ (thus $\gamma=0.33$), 
and four levels, giving 256 points, from which we randomly select the 
desired number. 

\begin{figure*}[htb!]
\centering
\includegraphics[width=0.48\textwidth]{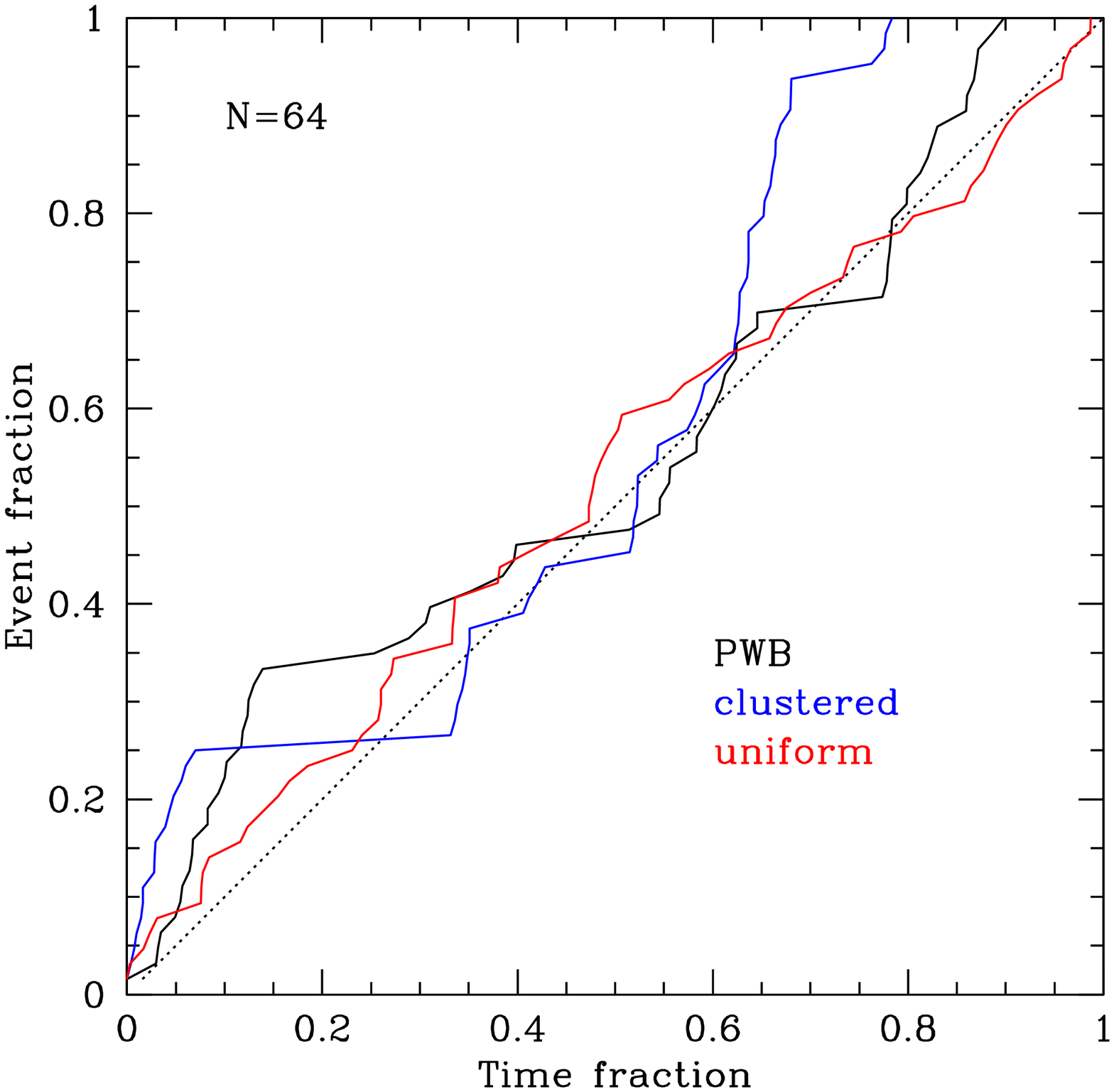}
\includegraphics[width=0.48\textwidth]{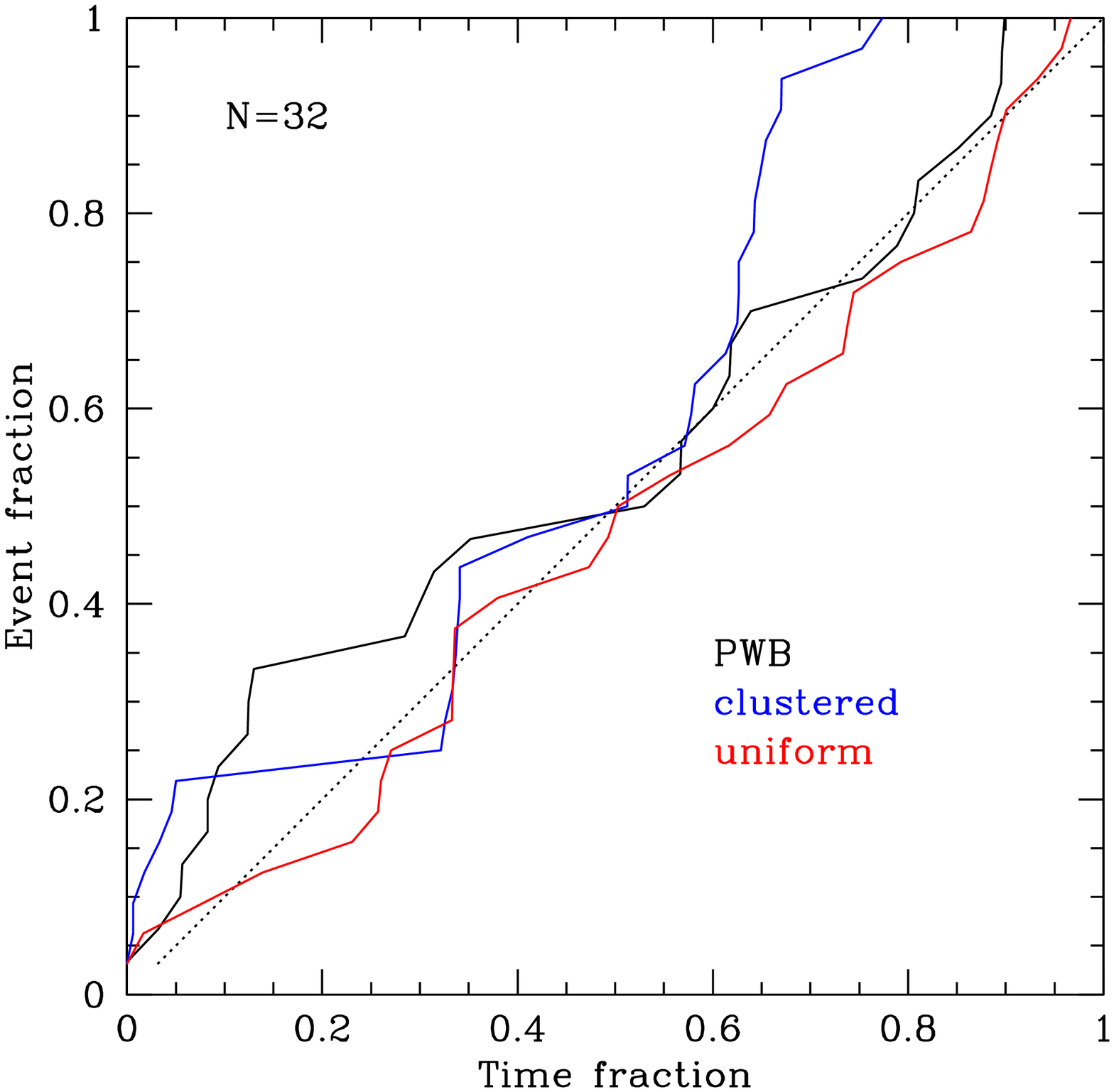}
\caption{As Figure~\ref{fig:pwball128} but for 64 (left) and 32 (right) mean 
events in each distribution. Random scatter increases the difficulty of 
discerning periodic windowed behavior below $N=64$. 
}
\label{fig:pwball3264}
\end{figure*}

For the PWB distribution we take four periods of length 0.25 with 
active fraction 60\%, i.e.\ an activity length of 0.15 and a gap length 
of 0.1. Within the active window the events are uniformly randomly 
distributed. While by eye one can discern gaps of no activity in the 
PWB CDF, 
one can as well in the clustered case, as well as lesser 
ones in the Weibull and even uniform distributions. As the data become 
sparser (and gaps become longer and realization scatter increases in the 
active windows), it is harder to assess visually whether there is actual 
PWB. 
Figure~\ref{fig:pwball3264} illustrates this as we reduce the number of 
events\footnote{Due 
to the random realization, 
in the PWB case while there are a mean of $N_{\rm total}/4$ events 
in each activity window, this number varies and so does the total. 
The three cases for PWB actually have 127, 63, and 30 events. The 
Weibull distribution case also has 125 rather than 128 events; since 
the Weibull distribution is well separated from the others, and to 
enhance clarity, we do not 
show it for the $N_{\rm total}=64$ and 32 
plots.} to 
$N=64$ and then 32.

Since the eye is suspect in determining PWB, we aim to quantify the 
difference in distributions more rigorously. One common approach to 
distinguishing between distributions is the relative entropy, or 
Kullback-Leibler (KL) divergence \cite{kl}. For two distributions $p(x)$ 
and $q(x)$ this is defined as  
\be 
D_{\rm KL}\left(p(x),q(x)\right)=\sum_i p(x_i)\,\ln\frac{p(x_i)}{q(x_i)}\,. 
\ee 
Generally one does not compare two realized distributions, since they 
may not have events at the same times $x_i$, but rather a realized 
distribution $p(x_i)$ is the data and $q(x)$ is the test or model 
distribution. 

We can now start to see why KL divergence is not ideal for 
our purposes: we cannot take $q(x)$, i.e.\ the model, to be the PWB 
distribution, since the important presence of no-event gaps gives $q(x)=0$. 
There are ways of getting around this using the CDF $Q(x)=\int dx\,q(x)$ 
rather than the probability density $q(x)$ itself \cite{klcdf}, but the results 
are not wholly satisfactory. Briefly, we do find the ability to distinguish 
each of the distributions from the uniform random case, and from each 
other. However, the quantification of the degree of difference is not easily 
interpreted, and the information of the gaps -- the dog not barking -- is 
diluted. (Note that when $p(x_i)=0$ there is no contribution to the KL 
divergence, regardless of the model distribution $q(x)$). 

Therefore, we adapt a method used in galaxy spatial clustering, where the 
cosmic web of structure -- both connectivity and void regions -- carries 
important information. The friends of friends (FOF) method \cite{fof1,fof2} defines 
clusters of activity where neighbors lie within a linking length $b$. For our one 
dimensional time series, this is trivial to implement: $t_i-t_{i-1}\le b$. 
This is fast and easy to apply to all our distributions. We set the 
linking length several times larger than the 
average uniform separation so that clusters will not 
be (rarely at least) falsely identified in 
uniform distributions. 
For example, $b=5/\bar n$, where $\bar n$ is the 
number of events divided by the time interval, 
i.e.\ the mean density or reciprocal of the inter-event 
spacing for a uniform distribution, 
roughly corresponds to a signal to noise $S/N\approx5$ distinction from a uniform  
distribution. Empirically, we find this works well. 

Friends of friends, like any clustering method, will have 
difficulties if there are few data points, but note that 
even if there is only one event in a window, FOF will 
recognize it as its own cluster unless $\bar n$ is too small. 
In any case, we will end up using FOF as a means of identifying 
windows, and turn in Sec.~\ref{sec:det} to other statistical 
techniques for robust quantification of  their characteristics. 
In the end, we will employ FOF quantitatively simply as a 
useful guide to reasonable priors for a detailed estimation 
procedure. 

An immediate output of applying the FOF method 
are the values from each activity cluster  found 
of the 
length of activity windows $a_j$, of gaps $g_j$ 
between them, and the possible periods 
from summing consecutive active and inactive times $T_j=a_j+g_j$ and the active 
fractions or duty cycles $f_j=a_j/T_j$. For true PWB we expect consistency (i.e.\ 
a narrow distribution) in $a_j$, $g_j$, etc.\ while these would be widely 
scattered or sparse for event distributions without some periodic 
behavior. 

Since PWB events occur {\it somewhere\/} in an active window, not 
necessarily at the extremes, we expect the measured values of $a_j$ to 
give lower limits on the true value $a$, measured values of $g_j$ to give  
upper limits on the true $g$, and the period $a_j+g_j$ to have some scatter 
around the true period $T$. All estimates become more accurate 
with more events in the time series; Section~\ref{sec:det} discusses 
accurate characterization of the PWB. 

\begin{figure}[htb!]
\centering
\includegraphics[width=\columnwidth]{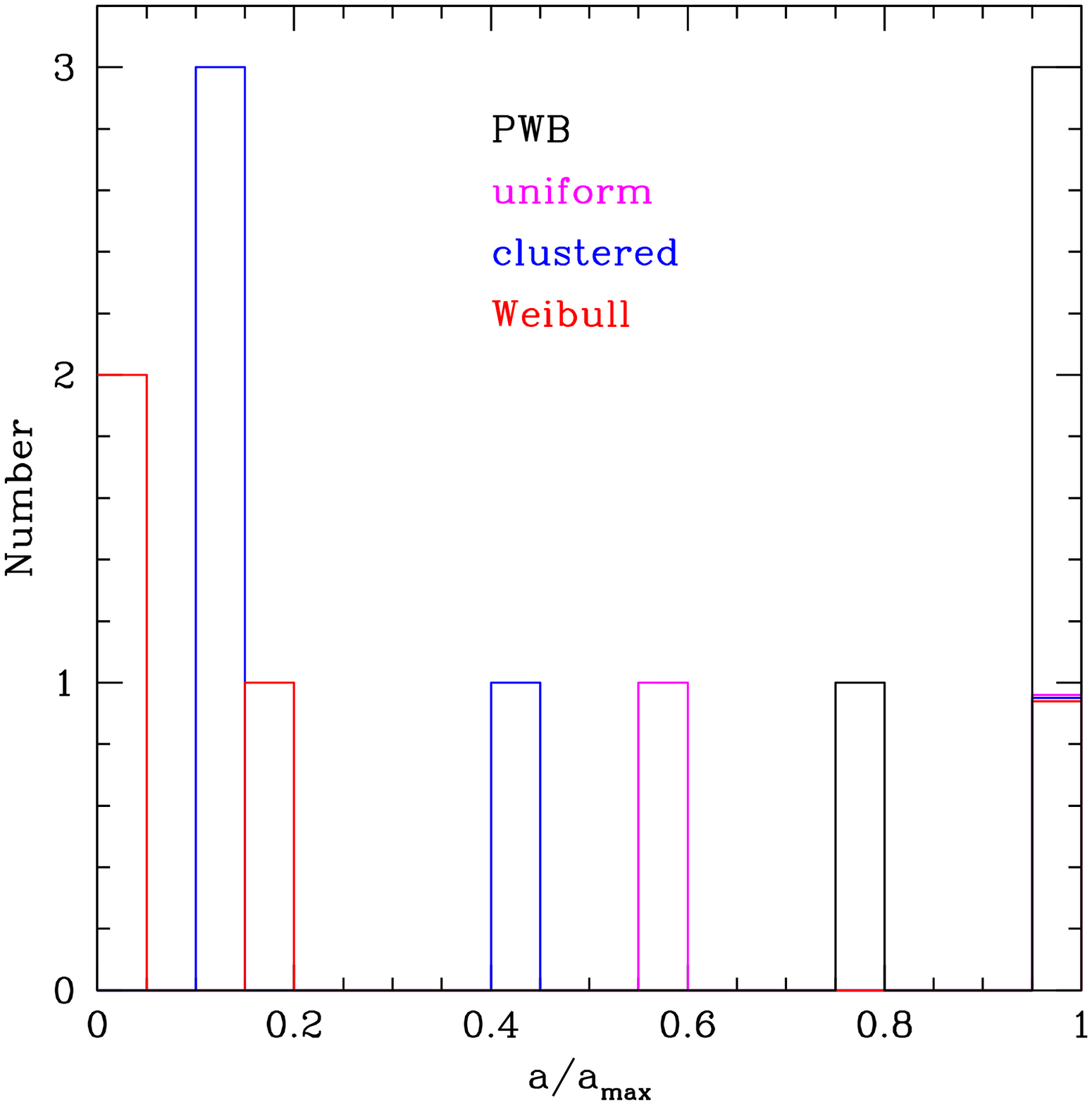} 
\includegraphics[width=\columnwidth]{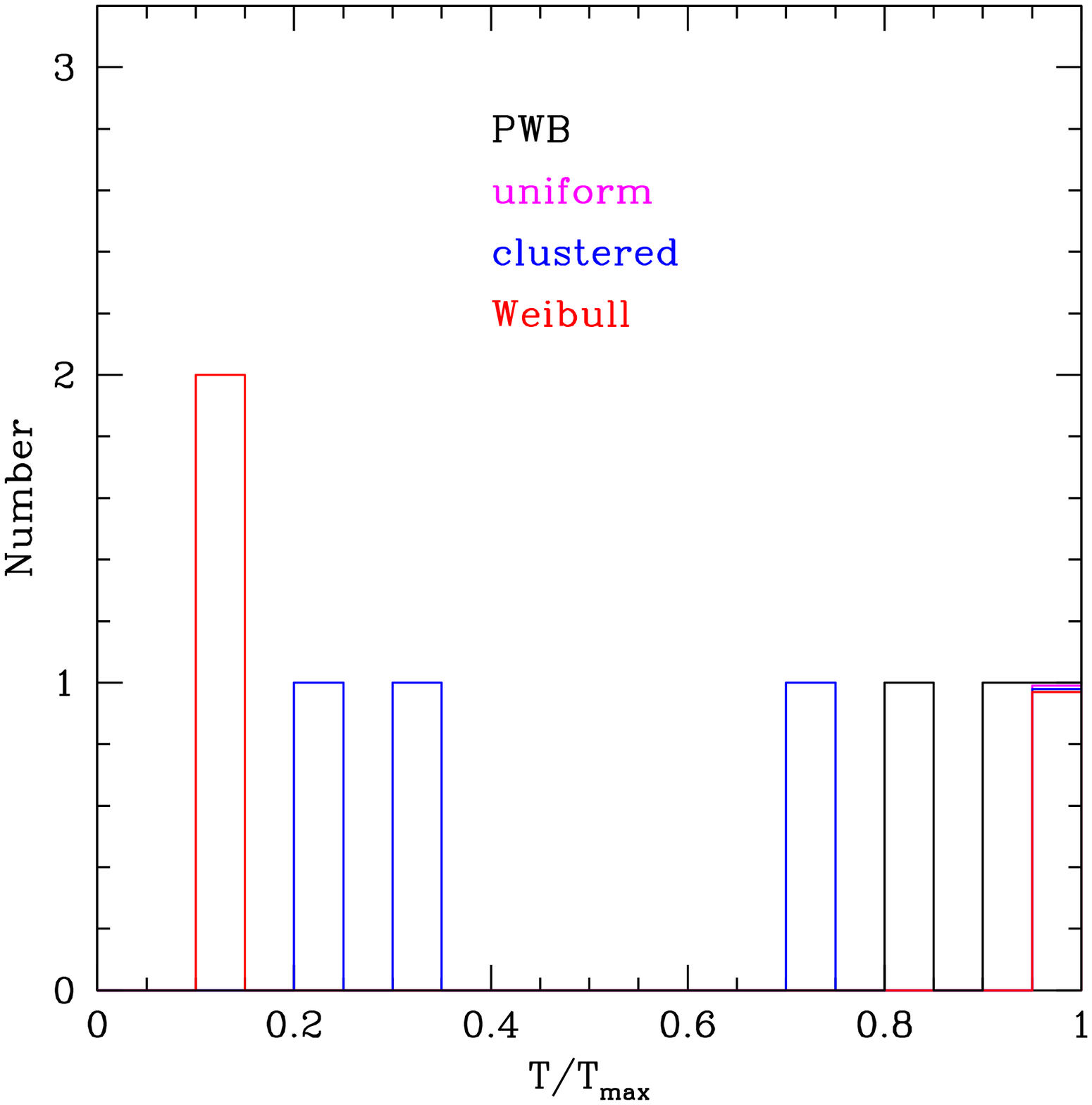} 
\caption{Histograms of the activity window 
durations (top) and periods (bottom) are shown for 
four realized distributions. Peaks, 
or narrow distributions, in both indicate periodic windowed 
behavior. To compare distributions more evenly we present 
the variables 
as fractions relative to the maximum active window duration and 
period for that distribution. (Thus all distributions 
by definition have one element for each quantity at the value 1, and we 
offset these slightly vertically for visibility; for the uniform case, e.g., 
the other instance is at $a/a_{\rm max}=0.34/0.60=0.57$.) Only the true PWB  
distribution shows a narrow distribution around 1 in the period, as well 
as in the active window duration. 
}
\label{fig:histo}
\end{figure}

Figure~\ref{fig:histo} shows the results of the FOF clustering analysis 
on the four types of distributions realized. The uniform random distribution 
was found to have two, disparate clusters, one extending from the first 
event to 34\% of the observing duration, the other extending for the last 
60\%, showing no clear periodic windowing. The Weibull distribution has 
four clusters, with activity window lengths scattered from below 1\% to 
64\% of the duration, again no PWB evident. The clustered distribution  
has five clusters with three activity windows at 2\% length, but others 
at 7\% and 17\% length; the related ``periods'' (sums of consecutive 
active and gap states) are more diverse, from 8\% to 33\% length, 
so no PWB is falsely detected. 
To compare the different distributions more 
clearly, we normalize each by their maximum 
values of period and activity found, 
$T_{\rm max}$ and $a_{\rm max}$ respectively. 

Finally, analysis of the actual PWB distribution case has 
three activity windows of 15\% length and one of 11\% length, with 
periods grouped from 23\% to 27\%. Recall that the input for generating 
the random realization was activity 0.15 and period 0.25, so FOF 
successfully reconstructs the truth. One can quantify this by computing 
the mean and standard deviation of the activity window length, for example, 
to assess the peaked nature of the estimation 
distribution\footnote{We 
use a weighted mean, 
\be 
\bar a=\frac{\sum_j N_j a_j}{\sum_j N_j}\,, 
\ee 
where $N_j$ is the number of events within the activity window of 
duration $a_j$. To the extent the scatter goes as $1/\sqrt{N_j}$, this 
is inverse variance weighting. Using this mean, we calculate the standard 
deviation.}. 

We go into characterization of PWB 
properties in more detail in the next section, going beyond FOF. Here we have motivated 
that PWB can be recognized, and that other distributions, even those 
that have innate clustering (Weibull or the correlation function clustered 
cases) do not lead to false PWB identification.

\section{Measuring Periodic Window Properties} \label{sec:det} 

Now that we have some confidence that PWB can be identified accurately, 
we turn to robust characterization of PWB properties such as the 
period and activity fraction.

\subsection{Quick look} 

We begin with a quick look estimate that we will find is 
surprisingly accurate, and useful to set a prior range 
for the more robust determination in the next subsection. 
As mentioned, measurements $a_j$ of the  
active window length give lower bounds for $a$, and those of gaps $g_j$ give 
upper bounds for $g$. A zeroth order estimate of the true quantities could then 
be simply the highest and lowest values, respectively. However, we 
would like to do better since we know the true $a\ge {\rm max}\ a_j$, plus 
we would like some indication of the uncertainty range. 

We improve the estimation by looking at the difference between the 
highest $a_j$, call it $a_{\rm hi}$ and the next two closest values, call 
them $\Delta a_1$ and $\Delta a_2$. If $\Delta a_2>2\Delta a_1$ we take 
$\Delta a=\Delta a_2$, otherwise $\Delta a=\Delta a_1$. For the gap 
length, this is more subject to overestimation (i.e.\ it is easier to 
not have a burst, even in an active window) so we start from 
the lowest $g_j$ and go down by $\Delta g=2\Delta g_1$ always. 
Then our estimate of the activity window 
length is $a=[a_{\rm hi},\,a_{\rm hi}+\Delta a]$ and the gap length is 
$g=[g_{\rm lo}-\Delta g,\,g_{\rm lo}]$. The estimate of the period is 
$T=[a_{\rm hi}+g_{\rm lo}-\Delta g,\,a_{\rm hi}+\Delta a+g_{\rm lo}]$. 
Again, we note that we will employ these FOF estimates 
simply as a useful guide to reasonable priors for a detailed 
estimation procedure. Nevertheless, we find below that they 
work quite well in the tests made. 

For our PWB with four windows, we find $a=[0.1486,0.1509]$, 
$g=[0.07367,0.1019]$, $T=[0.2222,0.2528]$, where the truth values are 
0.15, 0.1, 0.25. 
Note that for the PWB cases with $N=63$ and  $N=30$ 
realized events, the active window lengths and periods are 
still determined, but in the $N=63$ case the true period lies 
slightly outside the estimated range (at 1.6 times the mean 
uncertainty from the central value), and in the $N=30$ case 
the period has a 25\% uncertainty. Thus, estimation becomes 
more robust with $\gtrsim100$ events (at least for observations 
limited to four windows). 

As a blind test, one of the authors generated a distribution with 128  events and 
another fit it, obtaining $a=[0.0362,0.0398]$, $g=[0.0854,0.0885]$, 
$T=[0.1216,0.1283]$, and deducing correctly there were eight activity 
windows. The truths were revealed to be 0.0375, 0.0875, 0.125. 
A more difficult blind test used a distribution where some of the activity 
windows were empty, i.e.\ appeared as gaps rather than active times. 
Here the fits gave $a=[0.0374,0.0394]$, $g=[0.0856,0.0968]$, 
$T=[0.1230,0.1362]$, and deduced correctly that while there were eight 
activity windows during the observing time range, only six exhibited 
bursts and one of those had only a  single burst. The truths were 
0.0375, 0.0875, 0.125. 

Thus, in all cases we correctly reconstruct the PWB characteristics. 
We would, however, like to reduce the uncertainties further (the blind 
tests obtained the periods with 2.7\% and 5.1\% uncertainty respectively). 
This can 
be done with more data, of course, e.g.\ more bursts within an active 
window, a larger activity fraction (the blind cases had only 30\% duty 
cycle), or a longer observing duration giving more windows. Since we 
cannot control the first two, astrophysical properties, and we do not 
always want to wait for the last one, we instead use the first round 
of results as input to a more rigorous likelihood optimization routine. 
The initial estimates serve to guide priors that increase the 
speed and efficiency of the likelihood code.

\subsection{Robust estimation} 

For more robust determination of the period and activity 
window fraction, 
we carry out a likelihood analysis through a direct 
parameter grid search. The grid search is the most accurate 
approach, and tractable due to the low dimensionality of the 
parameter space. Other sampling methods are less efficient 
due to the posterior surface actually being a broad plateau, 
not an isolated peak.  For example, for any given period, 
a 100\% active window fraction means that the entire 
observational duration is treated as active and this will 
fit the data as well (though much less efficiently) as isolated 
windows. 

Once the PWB nature of a burst time series is indicated, 
we employ the range derived in the FOF analysis as 
an efficient guide for the grid search, placing 
top hat 
priors on the period $T$ and active length $a$. 
We introduce a phase parameter $\tau$ as well 
to describe the difference between the starting time 
of the first active window and the first burst observed, 
with a range $[-a,0]$ (so the prior on $a$ fixes the 
prior on  $\tau$ as well; note that FOF does 
not use phase information.) 
We 
check that the final results are not affected by these priors, 
they merely serve to make the grid search more efficient. 

The log likelihood function compares the model 
$\{T,a/T,\tau/a\}$ and 
the data, 
having two terms, 
\be 
\log \mathcal{L} = A+E\,. 
\ee 
The acceptance term $A$ is a step function, 
assigning zero if the data indeed falls within the 
activity windows of the model, i.e.\ the model 
describes the data. 
All models that fit the data, i.e.\ where burst events fall 
within an active window, have equal likelihood. However, 
if bursts 
fall into model gaps, where no events were predicted, 
a step to a large, constant negative penalty is assigned, 
preventing acceptance of the model (the exact size 
of the penalty does not matter if it is large enough,  
e.g.\ $<-1$). 
As mentioned above, this gives a broad plateau in the likelihood 
that allows trivially inefficient models, e.g.\ with $a=T$ 
and so having negligible or no window gaps. To break this 
degeneracy we add an efficiency term $E$ to the log 
likelihood that penalizes 
values of $a/T$ larger than necessary. For any given $T$ there 
will be a minimum (optimum) $a/T$, and the minimum $a/T$ across 
all $T$ defines the global optimum model $\{T,a/T\}$. 
The $E$ term serves to ``tilt'' 
the plateau so the optimization traces out its 
boundary. 
We use the form 
\be 
E=-\frac{a/T}{a_{\rm max}/T_{\rm min}}\,.  
\ee 
The numerator imposes a penalty for $a/T$ larger 
than strictly necessary, and the denominator is  
simply a constant normalizing factor not affecting the 
shape of the log likelihood, where $a_{\rm max}$ and $T_{\rm min}$ are the upper and lower 
prior bounds respectively, so that $-1<E<0$. 

We test this approach against the two mock data sets of the 
previous subsection: each contains 128 events distributed over  eight windows of activity, with varying numbers of bursts in 
each window. The first data set, denoted as ``full'', has all 
eight activity windows with events: (17,13,7,15,15,21,23,17) uniform randomly distributed in the respective windows. The ``sporadic'' 
data set contains the same number of bursts, however, two 
activity windows are empty and another one contains a single 
burst -- (11,3,36,0,1,52,0,25) -- to mimic a different possible  
observational scenario (and one that we will see in the next 
section is closer to a particular actual data set). Recall that the 
FOF analysis was able to discern correctly the number of active 
windows in each case, and obtain estimates for $T$ and $a$. 

\begin{figure}[thb!]
\centering
\includegraphics[width=\columnwidth]{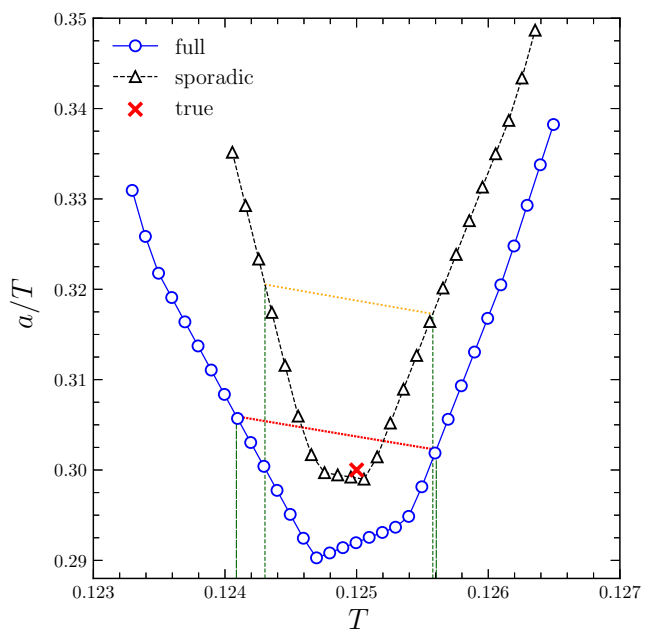}
\caption{
Likelihood optimization gives both local minima in active window fractions $a/T$ as a function of period $T$, and the 
global minimum. We show the cases for the two mock data 
samples, when all windows are active 
(``full''; solid blue curve) and when data is sporadic enough 
that some windows are empty of events (``sporadic''; dashed black 
curve). The global minimum is the best estimate; 
all points above the curves are consistent with the 
data, but past the dotted diagonal curves the models 
are inefficient (only a few are as restrictive as the 
data). We can use this region to define an uncertainty 
on $T$, shown by the projection to the $T$ axis. The input 
value is shown by the bold red x. 
}
\label{fig:opt8mis}
\end{figure}

Figure~\ref{fig:opt8mis} shows the minimum active fractions $(a/T)_{\rm min}$ found for each $T$, and the global minimum 
picking out the optimal $T_{\rm opt}$. 
From our likelihood analysis we obtain the best fit parameters 
to be $T=0.1248$ and $a=0.0362$ for the full, and 
$T=0.1251$ and $a=0.0374$ for sporadic mock cases respectively 
(compared to the truth, $T=0.125$, $a=0.0375$). Any model lying 
above the curve is a valid fit to the data, but less efficient 
than the optima. We see that the global optimum is quite close 
to the truth. Note that this method gives a best fit, but not 
an uncertainty per se. 

We can define an uncertainty by choosing to 
consider models with a bound on inefficiency 
such that the optimum behavior would not appear much 
less frequently than in 68.3\% of simulated data sets. 
For example, a model with a larger than needed active 
window, hence $a/T$, would be consistent with the 
data, but only rarely would its realizations 
be as restricted as the data, 
i.e.\ falling in narrower windows. 
``Efficient'' models lie in the region above the 
minimization curve but below the dotted, nearly 
diagonal inefficiency curve 
(see Appendix~\ref{apx:effic} for its expression). 
Projecting to the $T$ axis defines the 
range in the period. 
For the full case this gives a 
range of $T\in[0.1241,0.1256]$, or $[-0.5\%,+0.7\%]$ uncertainty  
on the period. For the sporadic case the range is 
$T\in[0.1243,0.1256]$, or $[-0.6\%,+0.4\%]$ uncertainty (the 
tighter precision relative to the full case is due to the higher density, 
though fewer, windows; 
recall they both have the same total number of events, 
just distributed differently).

\begin{figure}[htb!]
\centering
\includegraphics[width=\columnwidth]{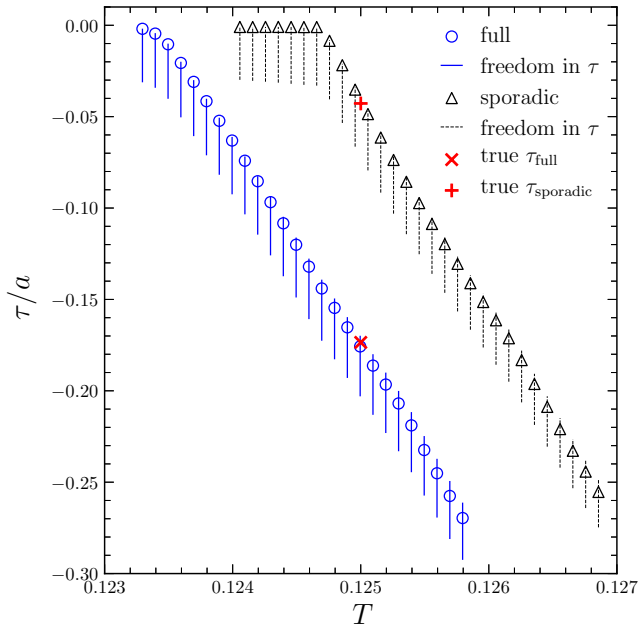}
\caption{
Likelihood optimization provides the phase fraction of the 
first observation, $\tau/a$, as a function of period $T$. 
For the minimum $a/T$ (the curves in Fig.~\ref{fig:opt8mis}), 
the phases are tightly constrained ($\sim$pointlike), but have more freedom 
for larger allowed $a/T$;  the vertical solid and dashed 
lines indicate the allowed phase range for $a/T=(a/T)_{\rm min}+0.01$. 
}
\label{fig:p8mis}
\end{figure}

Figure~\ref{fig:p8mis} shows the phase parameter $\tau/a$. It 
is tightly constrained when $a/T$ is at the minimum, but has a 
modest range when $a/T=(a/T)_{\rm min}+0.01$, for a given $T$. 
Finally, we note that 
we crosschecked against two other optimization approaches, 
specifically simulated annealing by employing the {\tt dual\_annealing} routine from the SciPy optimization package and a MCMC sampler, and found the results 
are compatible (see Appendix~\ref{sec:anneal}).

\section{Application to SGR1935+2154} \label{sec:sgr} 

Real observational data can be more difficult, and 
scientifically rewarding. We apply our methods to actual 
measurements of the source SGR1935+2154, testing whether 
it exhibits PWB, and giving robust estimates of its period and active window 
fraction. These characteristics can constrain models of 
the origin of the emission and properties of the system. 

SGR1935+2154, a magnetar within our Galaxy, was identified as the source of two FRBs occurring UT 2020 April 28 \cite{kirsten+20,bochenek20}. 
The local nature makes 
measurements of the object and its environment much easier and with much greater detail. In fact, a linkage between the production of soft gamma bursts in this source and FRBs is naturally suggested: the same two events as the FRBs, with the appropriate delay due to dispersion by interstellar electrons, were detected by $\gamma$-ray instruments \cite{zhang+20,Li+20,Mereghetti+20,Tavani+20,Ridnaia+20} (though some authors suggest that different source types may be responsible for these vs extragalactic FRBs \cite{margalit20}). 

As an example of the constraining power of these measurements, in \cite{grossan20} the SGR1935+2154 soft gamma burst PWB period was found to be 231 days, but the binary comb model for 
FRB \cite{iokazhang20} is limited to periods $\lesssim100$ days 
\cite{zhanggao20}. Such a model may perhaps be eliminated for SGR1935+2154 or other PWB  long-period magnetars \cite{grossan20},  
pointing to other models such as isolated neutron star precession 
(e.g.\ \cite{akgunlw06}) as the source of periodicity, 
coupled with a nonperiodic emission mechanism. 
More generally, given an excellent knowledge of the PWB parameters, correlating additional observational properties with these parameters may shed further light on the physical mechanisms at play. For example, with robust knowledge of the phase of the window boundaries, one could look not only for correlations with the modulation period, but also for effects at the window boundaries in SGR burst intensity, fluence, and spectral characteristics. Comparison of the variation in soft gamma burst  polarization measurements (coming from the next generation of instruments) with the PWB period and the window boundaries could shed light on the role of the magnetic field in these modulations, as well as in the role of magnetic field orientation in the emission mechanism. 

Long term monitoring of SGR1935+2154 has provided several 
years of data from a number of $\gamma$-ray instruments. 
In \cite{grossan20}, for some range of periods the data were folded at various trial periods, and the activity fraction -- the fraction of a period that would be consistent with all event data, was calculated, and the period with the minimum activity fraction was taken to be the ``best'' period. That analysis did not provide an examination of the uncertainty in the period or active fraction. 
Its conclusion that there was actually a periodic windowed behavior arose from comparison to uniform random events. 
Here we use different methods and quantification, and 
a clustering analysis. 

We apply our methods to burst data from the Third Interplanetary Network (IPN3 \cite{ipn}). IPN3 includes numerous spacecraft with X-ray and gamma-ray sensitive instruments, but notably the Konus instrument on the Wind spacecraft.  This instrument is in orbit around the sun at Lagrange point 1, far from earth, and so provides a nearly continuous, unobstructed view of the entire sky, and a more constant background than for low-earth orbit instruments. These are ideal properties for time series  
monitoring\footnote{While 
there is some heterogeneity in instruments and coverage, 
the main instruments other than Konus are in low earth  
orbit, without any long-term changing viewing zones and 
hence no bias for or against periods in the hundreds of 
days range. An analysis for particular sets of instruments 
was carried out in \cite{grossan20} and found results 
consistent with each other and the results here. 
}. 

\begin{figure}[htb!]
\centering
\includegraphics[width=\columnwidth]{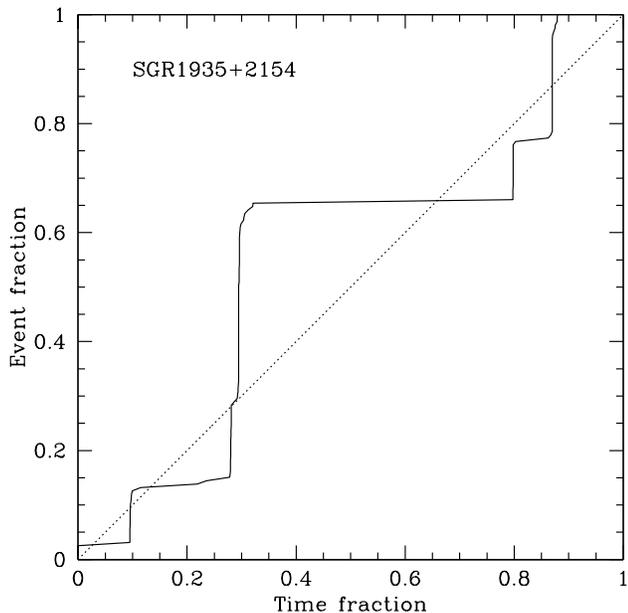}
\caption{
Cumulative distribution function for the SGR1935+2154 
data, similar to Fig.~\ref{fig:pwball128}. By eye it 
is difficult to confirm or deny periodic windowed  
behavior, requiring statistical analysis. 
}
\label{fig:cdfsgr}
\end{figure}

Carrying out our procedure for identifying and 
characterizing PWB, we 
first examine the CDF, as in Fig.~\ref{fig:pwball128}, 
but here applied to SGR1935+2154 data 
(159  events identified in the IPN3 SGR list for SGR1935+2154 as of 2021 February 1 \cite{ipn}). 
Figure~\ref{fig:cdfsgr} shows the resulting CDF. 
While there are  clearly episodes of activity and gaps of 
inactivity, it is difficult to tell by eye if there is 
PWB.

\begin{figure}[htb!]
\centering
\includegraphics[width=\columnwidth]{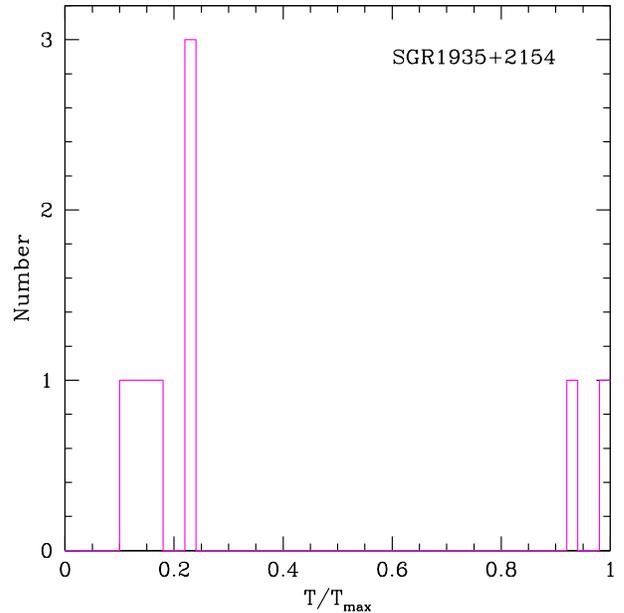}
\caption{
Histogram of the (pseudo)period, relative to the maximum 
instance, for the SGR1935+2154 data. The peak at 0.23 is 
suggestive, and the cluster of four just below may indicate the 
true period lies in between the two clusters, with some 
having sparse activity in the active window (hence 
apparently shorter periods) and some having long gaps (hence 
apparently longer periods). Quantitative analysis confirms 
this. 
}
\label{fig:histosgr}
\end{figure}

Applying the FOF method, we carry out the period analysis 
in Figure~\ref{fig:histosgr}. While the 
activity window lengths and gaps scatter greatly, the 
pseudoperiods (sum of consecutive active window lengths 
and gap lengths) show an interesting pattern. We exhibit 
$T/T_{\rm max}$, and see a concentration  
around $T/T_{\rm max}\sim0.2$, or $T\sim 0.1$.

The FOF method gives the estimate for the period 
(not pseudoperiod ratio) of $T=[0.056,0.129]$. While a broad 
estimation, due to some empty activity windows increasing 
the uncertainty, it still provides a useful prior for the  
more incisive likelihood analysis. In addition we find the 
activity window length $a=[0.049,0.077]$ and six active 
windows. 

\begin{figure}[htb!]
\centering
\includegraphics[width=\columnwidth]{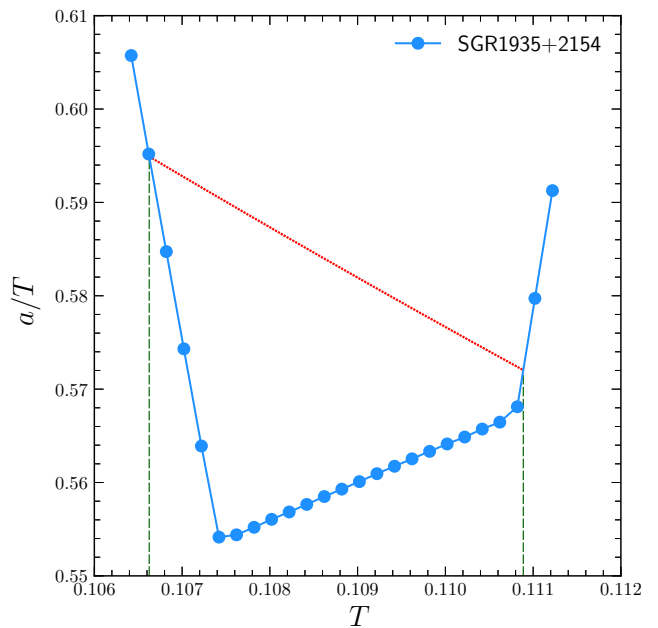} 
\caption{
As Figure~\ref{fig:opt8mis}, but for the actual SGR1935+2154  data. The global optimum gives $T=0.1074$, $a/T=0.554$. 
This corresponds to $T=230.6$ days, $a=127.8$ days.
}
\label{fig:optsgr}
\end{figure}

We proceed with our likelihood analysis and parameter determination 
of the SGR1935+2154 data in the same manner as the simulated cases 
of Sec.~\ref{sec:det}. The final results determine a global optimal 
period $T=0.1074$ and active fraction $a/T=0.554$ (i.e.\ $a=0.0595$). 
Converting back to days by rescaling to the duration of observations, 
this implies the PWB has period $T=230.6$ days, with an active 
fraction of 55.4\%. 
Figure~\ref{fig:optsgr} shows the results, along with the local optima 
boundary. 
Using the efficiency criterion, analysis of the SGR1935+2154 data yields a range of 
$T\in[0.1066,0.1109]$, 
or $[-0.7\%,+3.2\%]$ uncertainty on the period.

\begin{figure}[htb!]
\centering 
\includegraphics[width=\columnwidth]{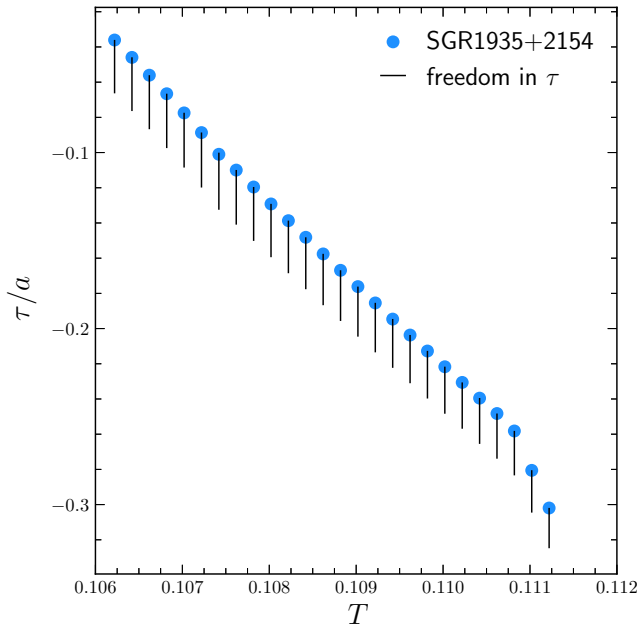} 
\caption{ 
As Figure~\ref{fig:p8mis}, but for the actual  SGR1935+2154 data. Here black vertical lines indicate the allowed phase 
range for $(a/T)_{\rm min}+0.02$. 
}
\label{fig:psgr}
\end{figure}

Figure~\ref{fig:psgr} presents the estimation of the phase parameter, 
with a best estimate of $\tau/a=-0.101$, i.e.\ the first detected 
burst of the time series occurred 10\% of the way through the 
activity window. The 
estimation of this is very tight ($-0.2\%,+0.9\%$) at 
$(a/T)_{\rm min}$, broadening as shown in the figure as one moves 
away from the local optima (here we show a shift by 0.02 from the minimum 
$a/T$, the same fractional difference as the 0.01 shift used in 
Fig.~\ref{fig:p8mis}).

Finally, we note that the FOF estimations for the 
period $T$ and active window length $a$ do include the 
likelihood optimization results of $T=0.1074$, $a/T=0.554$ 
(i.e.\ $a=0.0595)$, and the optimization results lie well 
inside the prior information from the FOF analysis. 
Our results agree as well with those from \cite{grossan20}.

\section{Conclusions} \label{sec:concl} 

Astrophysical bursts occur in observations throughout the 
electromagnetic spectrum, from the radio to optical to gamma ray. 
Repeated outbursts indicate the source is not totally disrupted, 
and periodic bursting points at some physics connected with, e.g., 
rotation or an orbital companion. An intriguing middle ground that 
is becoming more recognized with further data is periodic windowed 
behavior, where activity windows, rather than the burst events 
themselves, have periodicity. This can also provide important clues 
to the astrophysical mechanism of the burst, and system characteristics. 

Analysis methods for strict periodicity often fall short when dealing 
with PWB, as the duty cycle is important, activity windows can be 
empty of events, and the time series of events can be distributed 
in a complicated manner. We have emphasized that the lack of bursts 
carries critical information that must fold into the analysis, and 
we develop a time domain method that takes this into account. 

The cumulative distribution function of event intervals works well 
at identifying whether or not PWB is a reasonable possibility. We 
test this technique for four distributions: uniform random, Weibull, 
PWB, and a special distribution with correlated clustering. Given 
reasonable indications identifying PWB from the CDF analysis, we then 
draw on the friends of friends technique from galaxy clustering to 
characterize the PWB. This FOF analysis delivers quantitative 
estimates of the period $T$, active window length $a$ (and hence 
duty cycle $a/T$), and observing phase $\tau$. For our test cases of 
mock data, the estimates accurately reconstruct the input and have 
$\sim3-5\%$ precision for data with at least 100 events and $\gtrsim4$  
populated windows. However, we view the FOF analysis as a guide 
toward carrying out a full likelihood analysis (where the FOF can 
serve in setting reasonable priors). 

For the likelihood analysis we use an optimization grid approach, 
due to the low dimensionality of the parameter space, increased 
accuracy, and that 
the posterior surface is actually a broad plateau, not an isolated 
peak. However we do find consistent results with both a simulated 
annealing approach and a Markov Chain Monte Carlo approach. We 
minimize the duty cycle that agrees with the data, and find 
accurate estimates of the period to  $\lesssim1\%$ uncertainty on 
the mock data. Analyzing real observational data on the source 
SGR1935+2154 we first identify that PWB is reasonable for the data, 
and then characterize it as having period $T=230.6$ days, with $\sim1.9\%$ uncertainty, and duty 
cycle $a/T=55.4\%$. 

Finding PWB for what is truly a random distribution is 
highly unlikely: if we consider the probability that a 
uniform distribution realization would avoid all the 
10 spans of time that PWB predicts no activity (let 
alone have the active windows in a periodic pattern), 
this is $P=(1-0.554)^{10}\approx3\times 10^{-4}$. 
For a truly uniformly random distribution one could 
take into account not just the number of windows 
but the total number of events, so the probability 
would be $P\approx  (1-a)^N\approx 2^{-159}\ll1$. Nevertheless, 
the ultimate proof will be predictivity: if the 
values for the period and active fraction derived 
above are correct, the next two active windows 
(which admittedly are not guaranteed to have activity) 
are from
June 1 - October 7, 2021 and January 18 - May 26, 2022 and we predict no  activity outside of our active 
windows\footnote{On June 24, 2021 a burst was detected by the Fermi 
Gamma-Ray Space Telescope, in our predicted window. Equally importantly, 
no bursts were detected outside our predicted windows.}. 

Numerous next generation time domain surveys in a wide range of wavelength bands 
(e.g.\ LSST \cite{lsst} in the optical, DSA-2000 \cite{ds2000} and CMB-S4 \cite{cmbs4} in the 
radio and submillimeter, 
wide-field instruments such as STROBE-X WFM \cite{strobex} and SVOM ECLAIRs \cite{svom} in the X-ray/gamma ray bands) will greatly increase the database and diversity 
of repeating sources with possible PWB. The efficient methods 
presented here give a straightforward path for analysis, identification as PWB (vs, e.g., simply clustering), 
and its characterization. Accurate estimations of PWB, and the period, 
duty cycle, and phase, offer the potential for significant advances in 
understanding the physics of energetic bursts and the properties of 
the repeating outburst systems, in a wide variety of astrophysical 
contexts.

\acknowledgments 

We thank Alex Kim for helpful discussions. 
This work is supported in part by the Energetic Cosmos 
Laboratory. 
EL is supported in part by the 
U.S.\ Department of Energy, Office of Science, Office of High Energy 
Physics, under contract no.~DE-AC02-05CH11231. This paper has made use of data of 
the Interplanetary Network (\url{http://ssl.berkeley.edu/ipn3}), maintained by K.\ Hurley.

\appendix 

\section{Estimating Inefficiency}  \label{apx:effic} 

Many models, i.e.\ combinations of periods $T$ and active 
fractions $a/T$, can fit the data. Trivially, a model 
that is always active, $a/T=1$, will fit the burst data 
but be inefficient at doing so. That is, it allows bursts 
at any time but they seem to appear only within periodic 
windows. Such a formal fit is not informative. We therefore 
seek the most efficient fit as the most informative: the 
model with $(a/T)_{\rm min}$ and its associated 
$T_{\rm best}$. However, models close to this are only a 
little inefficient, that is many realizations of such 
models would generate data still falling within the 
optimal window structure. Conversely, for models further 
away, such as all-active $a/T=1$ models, Monte Carlo 
simulations of 
such a model would show many instances that, while 
including the data, would also have predicted many 
bursts where none were seen. Hence it is inefficient 
(or, if you like, complete but not pure). 

We seek a measure of the inefficiency, so that the 
region of efficient models can translate to a range, 
or uncertainty, of the period and active fraction. 
Consider a single window. If it is a little wider 
than optimal, then that model will fit the more 
restrictive data, but be somewhat inefficient at 
doing so. Suppose we want 68.3\% of simulations of 
a model to not only match the data but also not 
give bursts outside the optimal 
windows. For one window, and one burst within the 
window, this means that models with 
$a'/T'>(a/T)_{\rm min}/0.683$ are likely to be 
inefficient. For $N$ independent windows the 
inefficiencies $(a'/T')/(a/T)_{\rm min}$ multiply, 
so to obtain an efficiency 
$0.683<1/[(a'/T')/(a/T)_{\rm min}]^N$ 
each window can 
only contribute a factor $0.683^{-1/N}$. 
If we consider more 
than one burst within a window, we have to understand 
the coherence between bursts before we can 
quantitatively evaluate this, but $0.683^{-1/N}$ gives 
an upper limit to the inefficiency so we stay with 
this. 

Similarly, if the period is taken to be longer than optimal, 
this can also be inefficient. For a constant $a/T$, 
a longer $T$ means a longer active window width $a$. 
Again this adds inefficiency to each window, giving 
a factor $T'/T$ for each window. There are further 
effects from the shift of the far and near sides of 
the windows, and the phase, but these contribute less 
when $T'-T\ll a$, $a'-a\ll a$. Under those conditions 
we simply multiply the two inefficiency factors to 
get to first order 
\be 
(a/T)_{\rm eff}=(a/T)_{\rm min}\,(T'/T)^{-1}\,0.683^{-1/N}\,.  
\ee 
We use this to define the diagonal inefficiency 
curves in Figs.~\ref{fig:opt8mis} and \ref{fig:optsgr}. 
We have checked this gives a reasonable approximation under 
the conditions stated by running a suite of 1000 Monte 
Carlo realizations.

\section{Comparison to Direct Optimization} \label{sec:anneal} 

As mentioned in Sec.~\ref{sec:det}, we have crosschecked 
our direct grid search optimization routine with a 
standard MCMC and a simulated annealing optimization. 
The direct grid search is innately more exact, and 
sufficiently efficient due to low dimensionality of our 
parameter space that we use it throughout the paper. 
We exhibit some results of the other two 
methods here. 

\begin{figure}[htb!]
\centering
\includegraphics[width=\columnwidth]{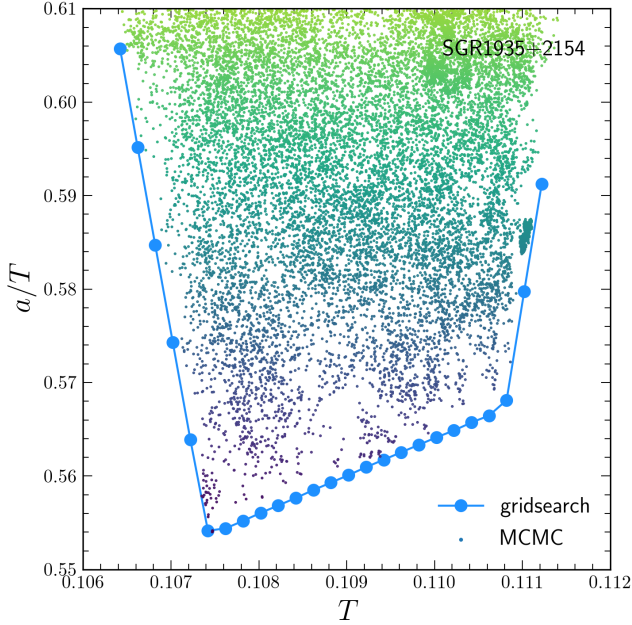} 
\caption{
MCMC samples plotted in the active window fraction $a/T$ vs period $T$ subspace of the full 3-dimensional parameter space (including the phase $\tau$). 
The color gradient shows the ``tilt'' from the efficiency term in the log likelihood. 
The MCMC optimal solutions for the 
local minima are bounded from below by the optimal values found in the grid search optimization.
}
\label{fig:mcmc}
\end{figure}

\begin{figure}[htb!]
\centering
\includegraphics[width=\columnwidth]{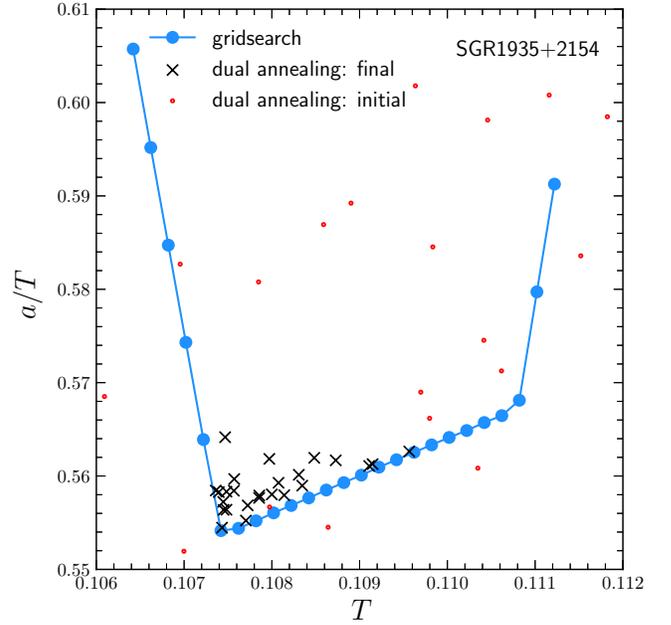} 
\caption{
Dual annealing likelihood optimization yields active window fractions $a/T$ very close to the global minimum found by the grid search method. Here, we show 25 randomly chosen initial points in the parameter space (red dots) 
and the final values (black x's). The final points represent the estimates of the 
global minimum (i.e.\ overall best fit rather than local minima), which can be refined further by repetitive application of the dual annealing approach.
}
\label{fig:anneal}
\end{figure}

Fig.~\ref{fig:mcmc} shows the MCMC samples generated using the Stan software \cite{stan} to find the global optimum model and explore the distribution of other combinations of parameters $\{T,a/T,\tau/a\}$ consistent with SGR1935+2154 data. 
It agrees well with our direct 
optimization. 
We also used the dual simulated annealing optimization routine \texttt{dual\_annealing} \cite{anneal} as a complementary approach to the grid search optimization to quickly find the best estimates of the global minimum. We randomly chose 25 initial points and found the dual annealing provided good estimates of the global minimum as well, as seen in Fig.~\ref{fig:anneal}.

\end{document}